\documentclass[twocolumn]{aastex631}

\usepackage{amsmath}

\received{\today}
\shorttitle{Am I loopy, or is there a planet in IM Lupi?}
\shortauthors{Verrios et al.}

\graphicspath{{./}{Images/}}
\usepackage{graphicx}
\usepackage{float}
\usepackage{amssymb}

\begin{document}

\title{Kinematic evidence for an embedded planet in the IM Lupi disc}

\author[0000-0003-3803-3855]{Harrison J. Verrios}
\affiliation{School of Physics and Astronomy, Monash University, Clayton, Vic 3800, Australia}

\author[0000-0002-4716-4235]{Daniel J. Price}
\affiliation{School of Physics and Astronomy, Monash University, Clayton, Vic 3800, Australia}

\correspondingauthor{Daniel J. Price}[]
\email{daniel.price@monash.edu}

\author[0000-0001-5907-5179]{Christophe Pinte}
\affiliation{School of Physics and Astronomy, Monash University, Clayton, Vic 3800, Australia}
\affiliation{Universit\'e Grenoble Alpes, CNRS, IPAG, F-38000 Grenoble, France}

\author[0000-0001-7641-5235]{Thomas Hilder}
\affiliation{School of Physics and Astronomy, Monash University, Clayton, Vic 3800, Australia}

\author[0000-0001-7764-3627]{Josh Calcino}
%\altaffiliation{jcalcino@lanl.gov}
\affiliation{Theoretical Division, Los Alamos National Laboratory, Los Alamos, NM 87545, USA}

\begin{abstract}
We test the hypothesis that an embedded giant planet in the IM Lupi protostellar disc can produce velocity kinks seen in CO line observations as well as the spiral arms seen in scattered light and continuum emission.  We inject planets into 3D hydrodynamics simulations of IM Lupi, generating synthetic observations using Monte Carlo radiative transfer. We find that an embedded planet of 2--3 M$_{\rm Jup}$ can reproduce non-Keplerian velocity perturbations, or `kinks', in the $^{12}$CO~J=2--1 channel maps. Such a planet can also explain the spiral arms seen in 1.25\,mm dust continuum emission and $1.6\mathrm{\mu m}$ scattered light images. We show that the wake of the planet can be traced in the observed peak velocity map, which appears to closely follow the morphology expected from our simulations and from analytic models of planet-disc interaction.
\end{abstract}


\section{Introduction}
\label{sec:introduction}
IM Lupi, a young star located 155.8$\pm$0.5 pc away \citep{KlionerS.A2018GDR2}, hosts a large and spectacular protoplanetary disc \citep{Panic:2009vr,Cleeves:2016ux,Avenhaus:2018tb,Pinte:2018uz}. Observations in the $^{12}$CO~J=2--1 spectral line from the Disk Substructures at High Angular Resolution Project (DSHARP) \citep{Huang:2018,Andrews:2020wh} showed localised deviations from Keplerian velocity in multiple velocity channels. More recently, the Molecules with ALMA at Planet-forming Scales Survey (MAPS) \citep{Oberg:2021vx} have confirmed these structures are present within the disc and are not observational artefacts. \citet{Pinte:2020vw} predicted that a massive planet located at $117$~au from the host star could be the cause of these deviations, referred to as \textit{kinks} (see definition in \citealt{Calcino:2021hp}). Detecting this planet using traditional observational methods, such as radial velocities, transits or direct imaging, is currently impossible as its orbit is too large and the planet is hidden within an optically thick disc of gas and dust \citep[e.g.][]{Pinte:2018uz}.

Scattered light images of IM Lupi \citep{Avenhaus:2018tb} provide further evidence for embedded planets. This spectacular image revealed spiral structures in the upper layer of the disc traced by scattered light from sub-micron grains. Additionally, observations from DSHARP in the 1.25 mm dust continuum show spiral arms and gaps in the midplane dust disc \citep{Andrews:2020wh}.

%%% REMOVE
%A recent approach to modelling these systems is to use an N-body simulator that is able to model the effects of a large number of particles over a long time-span. A smoothed particle hydrodynamics (SPH) simulator solves the equations of Lagrangian hydrodynamics to model densities of particles, with mass $m$, as a fluid with some velocity $v$ \citep{Price:2018tl}. Additionally, we can use a radiative transfer code to model the effects of light from the disc to a virtual observer using accurate chemical and physical properties. In this analysis, we perform simulations of the IM Lupi protoplanetary disc with specific dust properties and multiple planet masses.

IM Lupi is important because it is one of only a few discs that show large-scale spiral arms in mm-continuum emission. The other well-studied example is Elias 2-27 where there are two main hypotheses \citep{Meru:2017ji}: an embedded planet \citep{Meru:2017ji} or gravitational instability \citep{Forgan:2018if,Hall:2018,Veronesi:2021,Paneque-Carreno:2021}. The main prediction has been that gravitational instability will produce flocculent spiral structure and hence `kinks everywhere' \citep{Hall:2020}, whereas planets should produce more localised deviations from Keplerian motion \citep{Pinte:2020vw}. Gravitational instability also requires a massive disc, while embedded planets do not require this.

Our aim in this Letter is to investigate whether an embedded planet in the disc can explain the observed substructures in IM Lupi. We model discs using hydrodynamical simulations, creating synthetic observations using Monte Carlo radiative transfer. Comparing our models to observations enables us to constrain the mass and location of the planet. We also predict kinematic perturbations from the planet wake in the peak velocity map, which we confirm are present in the observational data. Our paper is organised as follows: We describe our methods and initial conditions in Section~\ref{sec:methods}, present our findings in Section~\ref{sec:results}, discuss the implications and limitations in Section~\ref{sec:discussion} and conclude in Section~\ref{sec:conclusion}

%This could increase the number of exoplanets discovered around young stars, in particular planets at larger orbits, and could help provide further evidence into the formation of planets and structures in protoplanetary discs.

%%%%%%%%%%%%%%%%%%%%%%%%%%%%%%%%%%%%%%%%%%%%
%%%%%%%%%%%%%%%%% METHODS %%%%%%%%%%%%%%%%%%

\begin{figure*}
    \centering
    \includegraphics[width=\textwidth]{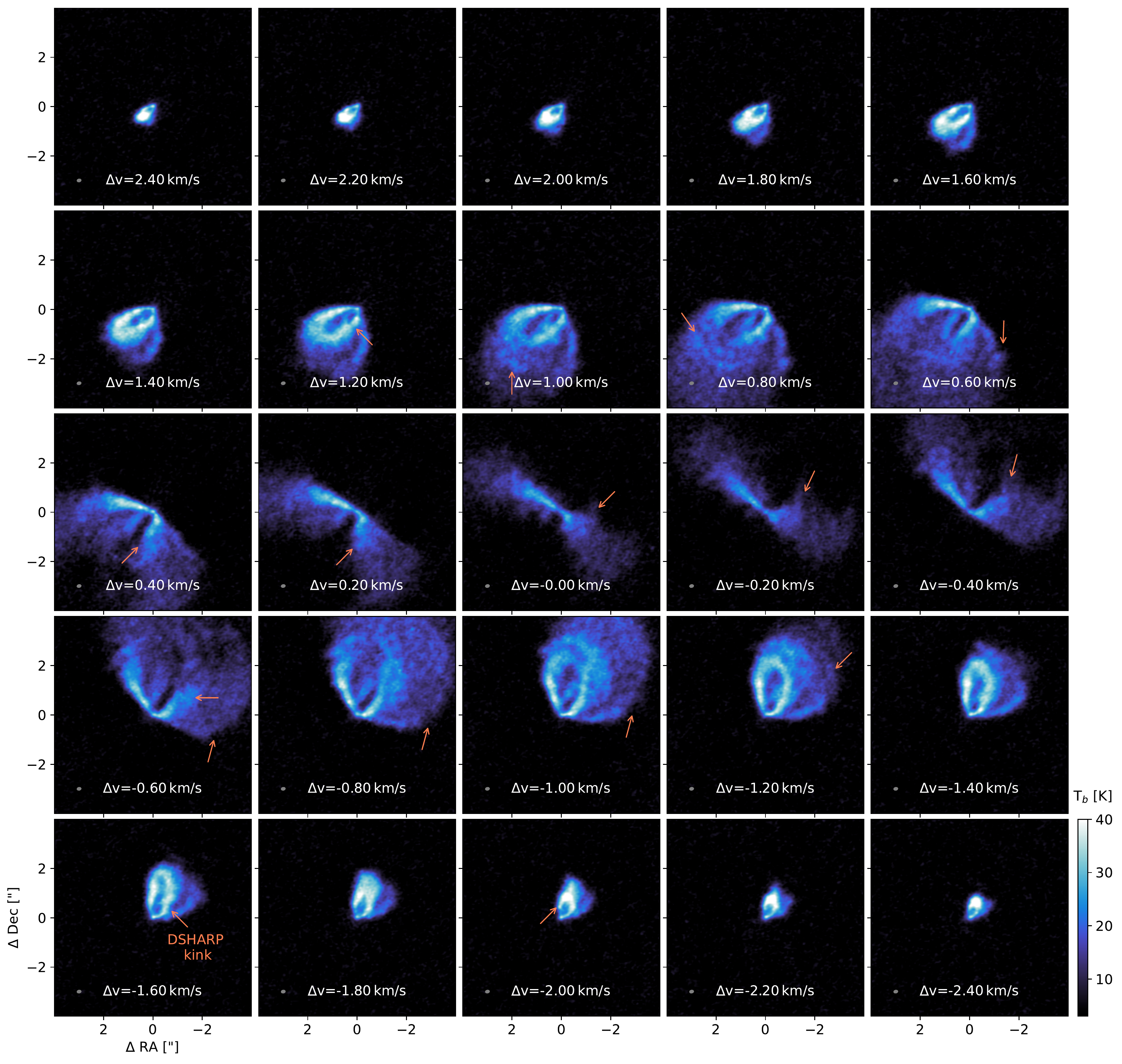}
    \caption{Channel maps of $^{12}$CO J=2--1 line emission in IM Lupi \citep{Oberg:2021vx}. Arrows identify non-Keplerian velocity perturbations visible in the data, including the tentative kink identified in the DSHARP data by \citet{Pinte:2020vw} (lower left panel). The channel maps for IM Lupi are available at the\dataset[MAPS website]{https://alma-maps.info/data.html}.}
    \label{fig:channels}
\end{figure*}

\begin{figure*}
    \centering
    \includegraphics[width=\textwidth]{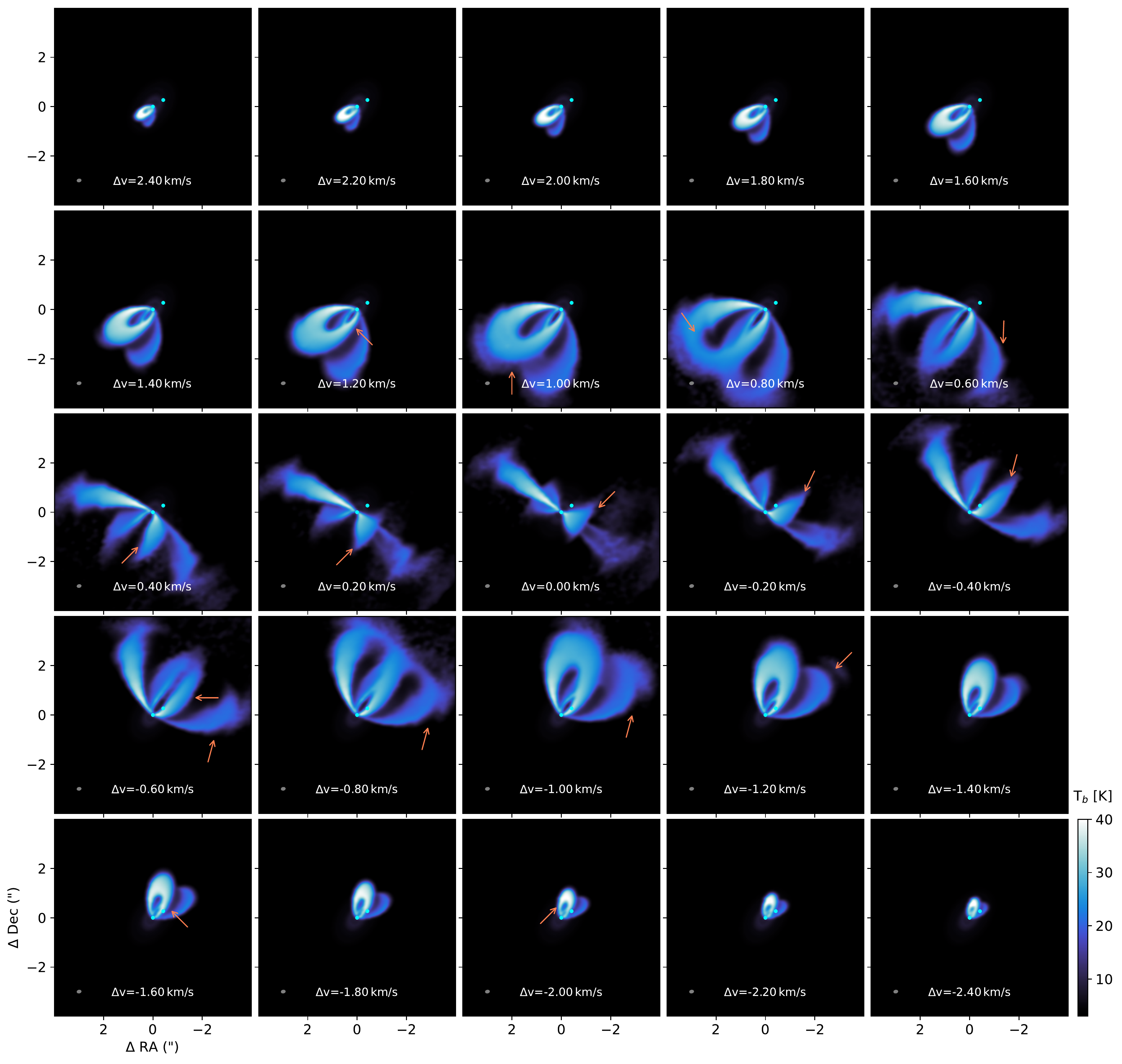}
    \caption{As in Figure~\ref{fig:channels}, but showing synthetic $^{12}$CO J=2--1 line emission from our 3 M$_{\rm Jup}$ simulation, post-processed with {\sc mcfost} and convolved with the observational beam. All of the non-Keplerian features labelled with arrows in Figure~\ref{fig:channels} find counterparts in our model, caused by the planet wake propagating through the disc. Our synthetic data cube of the IM Lupi protoplanetary disk is available on FigShare:\dataset[doi:10.26180/20145620.v3]{https://doi.org/10.26180/20145620.v3}.}
    \label{fig:channels_model}
\end{figure*}

\section{Methods}
\label{sec:methods}

\subsection{Initial Conditions}
\label{sec:met_parameterizing}

We performed smoothed particle hydrodynamics (SPH) simulations of planet-disc interaction using {\sc Phantom} \citep{Price:2018tl}. We modelled the disc with $10^7$ SPH particles, set up initially to follow
a tapered power-law surface density profile given by
\begin{equation}
\Sigma=\Sigma_c \left(\frac{R}{R_{\rm c}}\right)^{-p} \exp\left[-\left(\frac{R}{R_{\rm c}}\right)^{2-p} \right].
\end{equation}
We assumed $p=0.48$ \citep{Pinte:2018uz}. We used $R_\mathrm{c}=150$~au instead of the \citet{Pinte:2018uz} value of $R_\mathrm{c}=284$~au for computational convenience. We assumed a vertically isothermal equation of state $P=c_{\rm s}^2(R) \rho$ with $c_{\rm s} \propto R^{-q}$, $q=0.31$ and the sound speed normalised to give an aspect ratio $H/R=0.129$ at a radius of $100$~au \citep{Cleeves:2016ux,Pinte:2018uz}. We adopted a stellar mass of $1.12~\mathrm{M_\odot}$ \citep{Andrews:2020wh} with the star modelled as a sink particle with an accretion radius of 1~au. We setup the disc initially between $r_{\rm in}=30$~au and $r_{\rm out}=970$~au \citep{Panic:2009vr}, as we were not concerned with the inner disc structure. The vertically averaged ratio of smoothing length to disc scale height, $<h>/H$, varies between $\approx 1/20$ at 200 au and $1/5$ at 30 au, with around 10 resolution lengths per scale height at the final planet location and $\approx 5$ at the height of the CO and scattered-light emitting layer in IM Lupi \citep{Law:2021vy}. We adopted a total gas mass of $0.1~\mathrm{M}_{\odot}$, as determined by \citet{Cleeves:2016ux}. We also performed simulations with a disc mass of $0.01~\mathrm{M}_{\odot}$, but found that the higher disc mass better reproduces the scattered light image because the sub-micron sized grains remain well coupled in the top layers of the disc.

We included dust in the simulations using the {\sc multigrain} one-fluid algorithm \citep{Price:2015ab,Ballabio:2018tt,Hutchison:2018,Price:2018tl}. We modeled 11 grain sizes spanning $\mathrm{a_{min}}=1.0\mu\mathrm{m}$ to $\mathrm{a_{max}}=2300\mu\mathrm{m}$ on a logarithmically-spaced grid with a power law distribution with a slope of $-3.5$. We assumed a gas to dust ratio of 57 in order to give a total dust mass of $1.7~\times~10^{-3}$~M$_\odot$, as determined by \citet{Pinte:2018uz}.

%All disc parameters can be found in Table~\ref{tab:disc_parameters}.

%We evolved the simulations for a time corresponding to 66 orbits at the initial planet location.

%\begin{table}
%	\centering
%	\caption{{\sc Phantom} inputs used to simulate the IM Lupi protoplanetary gas and dust disc.}
%	\label{tab:disc_parameters}
%	\begin{tabular}{lrcl} % four columns, alignment for each
%		\hline
%		\textbf{Model Parameter} & \textbf{Value} & \textbf{Unit} & \textbf{Definition} \\
%		\hline
%%%		\hline
%		$\mathrm{r_{in}}$ & 30 & au & Disc Inner Radius \\
%		$\mathrm{r_{out}}$ & 970 & au & Disc Outer Radius \\
%		$\mathrm{r_{tap}}$ & 150 & au & Disc Taper Radius \\
%		$\mathrm{H/R}$ & 12.9 & & Scale Height \\
%		$\mathrm{M_{disc}}$ & 0.100 & $\mathrm{M_\odot}$ & Disc Gas Mass \\
%		$\mathrm{p}$ & 0.48 & & Viscosity Index \\
%		$\mathrm{q}$ & 0.31 & & Sound Speed Index \\
%		\hline
%		$\mathrm{dust:gas}$ & 0.0175 & & Dust to Gas Ratio \\
%		$\mathrm{a_{min}}$ & 1.0 & $\mu$m & Min. Grain Size \\
%		$\mathrm{a_{max}}$ & 2300 & $\mu$m & Max Grain Size \\
%		$\mathrm{n_p}$ & 10,000,000 &  & Particle Count \\
%		\hline
%	\end{tabular}
%\end{table}

\begin{figure*}
    \centering
    \includegraphics[width=\textwidth]{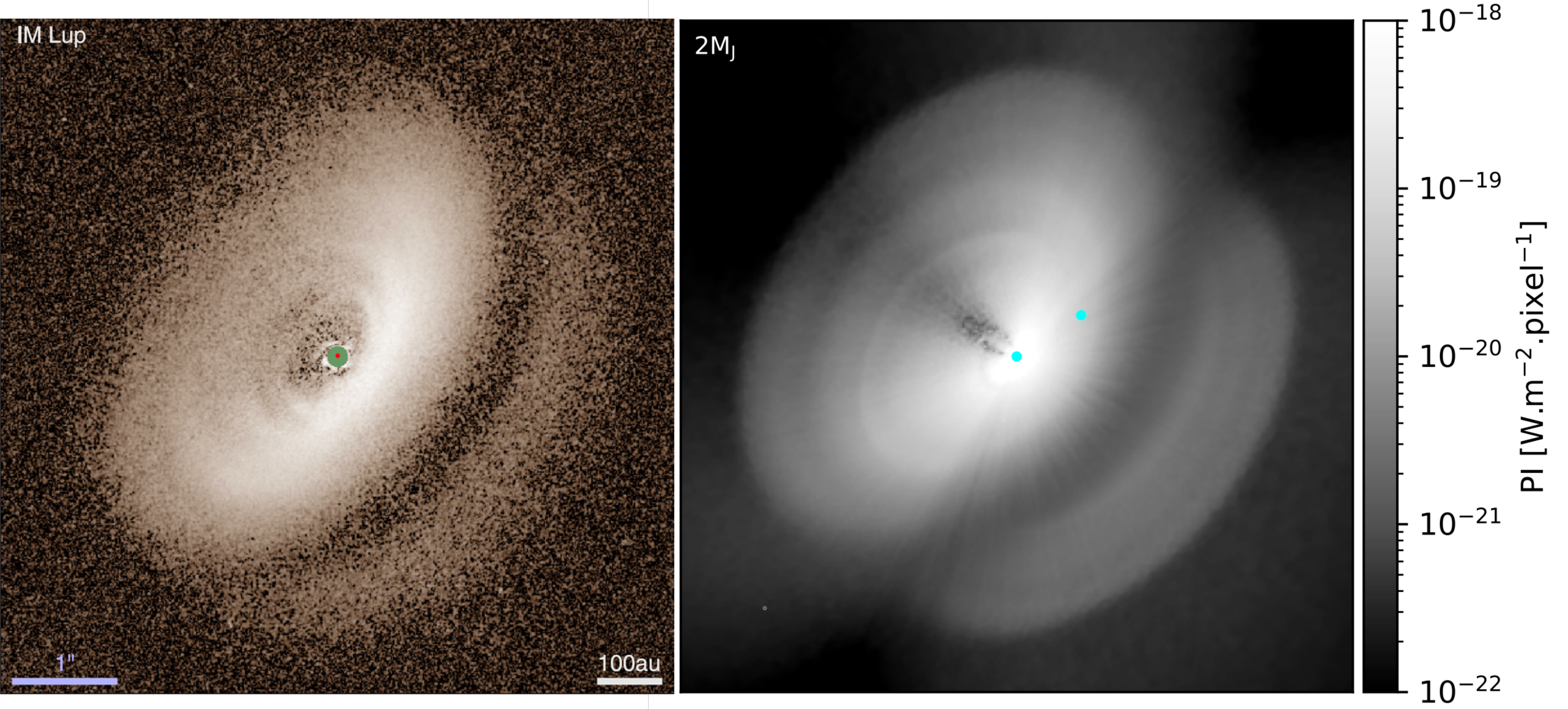}
    \caption{Scattered light. Left: VLT/SPHERE image from \citet{Avenhaus:2018tb} using Differential Polarimetric Imaging. Right: synthetic polarised intensity image at $1.6~\mu\mathrm{m}$ from our $2\mathrm{M_J}$ simulation. Panels are $6.4\arcsec \times 6.4\arcsec$. Cyan dots show positions of the star and planet, respectively. The planet wake in our model reproduces the spiral substructure observed in the upper layers of the disc. A fits file of our radiative transfer image is available on FigShare:\dataset[doi:10.26180/20145620.v3]{https://doi.org/10.26180/20145620.v3}
    %\cp{I don't like units per pixel as it depends on the pixel size, can we convert to per arcsec2 and/or normalize the image as the SPHERE data is not flux calibrated (or at least there is no info in the header)}
    }
    \label{fig:scattered_light}
\end{figure*}

\subsection{Embedded Planets}
\label{sec:met_embedded}

\citet{Pinte:2020vw} predicted a planet orbiting IM Lupi with a semi-major axis of 117~au, based on the location of the velocity kink in the $^{12}$CO channel in the DSHARP data. Since our planets migrate over the course of the simulation, we placed the planet initially at $150$~au, which results in a planet at approximately the correct radius. The initial fast migration (10~au/kyr) occurs mainly because it takes several orbits for the planet to open a gap. After this the migration steadies to $\approx 2.3$~au/kyr.

%However, the exact relationship between planet mass, disc properties and planet orbit is not presently known.

A consequence of the relatively high disc mass is that our planets, if modelled as `vacuum cleaner' sink particles would accrete a large amount of mass (up to ten times their initial mass in the first fifty orbits). Since this is largely an artefact of the boundary condition employed \citep{Ayliffe:2010ut}, we instead modelled our planet as a non-accreting sink particle with a softened gravitational potential, similar to the procedure in \citet{Szulagyi:2016}. We assumed a cubic B-spline softening kernel \citep[][]{Price:Monaghan2007} which becomes Keplerian at a radius $2h_{\rm soft}$, where $h_{\rm soft}$ is the softening length (see e.g. \citealt{Price:2018tl}). This allows the sink particle mass for the planet to remain constant throughout the simulation, while producing the correct interactions with the surrounding gas and dust. We use a constant softening length of $\mathrm{h_{soft}}=1.75$~au for the planet. To model the effects that the mass of the planet has on the disc, we performed four simulations, each with a single planet of mass 2, 3, 5 and 7 Jupiter-masses, respectively, plus one simulation with no planet. We kept all other initial conditions the same between each simulation.

\subsection{Radiative Transfer}
\label{sec:met_radiative}

%\begin{table}
%	\centering
%	\caption{{\sc Mcfost} parameters used to create synthetic observations.}
%	\label{tab:Mcfost_parameters}
%	\begin{tabular}{lrcl} % four columns, alignment for each
%		\hline
%		\textbf{Parameter} & \textbf{Value} & \textbf{Unit} & \textbf{Definition} \\
%		\hline
	 %   D & 156 & pc & Distance \\
	%	i & 230 & $^{\circ}$ & Disc Inclination \\
%		PA & 53 & $^{\circ}$ & Position Angle \\
%		\hline
%		$\mathrm{T_{star}}$ & 3900 & $^{\circ}\mathrm{K}$ & Star Temperature \\
%		$\mathrm{R_{star}}$ & 2.15 & $\mathrm{R_\odot}$ & Star Radius \\
%		\hline
%	    fluffy & 0.1 & $ $ & Grain Fluffiness \\
%		$\mathrm{M_{dust}}$ & 1.75$\times$ $10^{-3}$ & $\mathrm{M_\odot}$ & Dust Mass \\
	%	$\mathrm{f_{UV}}$ & 0.09 &  & Flux Density \\
	%	$v_{\mathrm{turb}}$ & 0.08 & km s$^{-1}$ & Turbulence Velocity \\
	%	$X\big(^{12}\mathrm{CO}\big)$ & 5 $\times$ 10$^{-6}$ &  & $^{12}$CO Abundance \\
	%	\hline
	%	$\mathrm{n_{grid}}$ & 401 & $\mathrm{px} $ & Grid Size \\
	%	$\mathrm{size}$ & 1000 & $\mathrm{au}$ & Imaging Size \\
	%	\hline
	%	$\mathrm{p_{az}}$ & 30 & $^{\circ} $ & Planet Azimuth \\
	%	$\lambda_\mathrm{cont}$ & 1250 & $\mu\mathrm{m}$ & Continuum Wavelength \\
	%	$\lambda_\mathrm{scatt}$ & 1.6 & $\mu\mathrm{m}$ & Scattered Wavelength \\
%		\hline
%	\end{tabular}
%\end{table}

We post-processed our 3D simulations using {\sc Mcfost} \citep{Pinte:2006uj,Pinte:2009vs}.
%The resulting continuum and velocity channel maps can be used to directly compare with the observational data.
We assumed the IM Lupi disc to be located $156$ pc from the observer \citep{KlionerS.A2018GDR2} with $230^\circ$ inclination (i.e. 50$^\circ$ but rotating clockwise) and a position angle of $53^\circ$ \citep{Avenhaus:2018tb}. We assumed $T_{\rm eff}=3900\mathrm{K}$ and $R_*=2.15\mathrm{R_\odot}$ \citep{Pinte:2018uz} for the central star.

We produced channel maps with a channel spacing of 0.05 km/s, from -2.4 km/s to +2.4 km/s.  For the channel maps we assumed turbulent broadening of $v_{\rm turb}=0.08$ km/s and $f_{\rm UV}=0.09$ \citep{Pinte:2018uz}. We adjusted the azimuthal location of the planet based on \citet{Pinte:2020vw}, giving an azimuthal angle of $30^\circ$ (clockwise from North). To avoid CO being too bright in the outer disc we adopted a slightly lower (fixed) $^{12}$CO-to-H$_2$ ratio of $1\times 10^{-5}$ compared to $5 \times 10^{-5}$ assumed previously \citep{Pinte:2018uz}. This is likely due to our reduced value of R$_{\rm c}$.

We assumed a wavelength of 1.25 mm to produce the dust continuum image \citep{HuangJane2018TDSa} and a wavelength of 1.6 microns to produce the scattered light image \citep{Avenhaus:2018tb}. For the scattered light comparison, we computed the polarized intensity (PI) image by combining the $Q$ and $U$ Stokes parameters \citep{Pinte:2006uj}. We assumed spherical grains (Mie scattering) to compute the polarised emission, as described in \citet{Pinte:2006uj}, assuming silicate grains \citep{Weingartner01} with grain sizes and distributions taken from the hydrodynamic model. To match both the continuum and scattered light images, we needed to rescale the grain sizes as in \citet{Pinte:2019tc}, scaling the notional Stokes numbers from the hydrodynamical simulations (down) by a factor of 10. This mainly suggests that grains are not spherical (see Section~\ref{sec:discussion}).

%%%%%%%%%%%%%%%%%%%%%%%%%%%%%%%%%%%%%%%%%%%%
%%%%%%%%%%%%%%%%% RESULTS %%%%%%%%%%%%%%%%%%

\section{Results}
\label{sec:results}

%We present four main results from our simulations, each indicating the effect embedded planets have on the observed disc morphology. These are: $^{12}$CO velocity channels, $\lambda=1250 \mu\mathrm{m}$ dust continuum, the $\lambda=1.6 \mu\mathrm{m}$ scattered light image and a peak velocity map showing the maximum gas velocity. The results were taken at identical simulation time, after $t=1.4 \times 10^4$ years.

\subsection{Kinks and non-Keplerian features in channel maps}
\label{sec:vel_velocity}

Figure~\ref{fig:channels} shows the observed $^{12}$CO~J=2--1 channel maps of the IM Lupi disc from the MAPS project \citep{Oberg:2021vx}. A typical Keplerian butterfly pattern is visible. Distinct CO-emitting surfaces from the top and bottom of the disc are also evident \citep[see][]{Pinte:2018uz}. On top of this butterfly pattern we indicate (with arrows) at 16 locations where `kinks' or other perturbations away Keplerian motion are visible in the data. This seemingly argues against a planetary origin since previous papers suggested that an embedded planet should produce only localised velocity perturbations \citep{Pinte:2018va,Pinte:2020vw}.

Figure~\ref{fig:channels_model} shows the synthetic channels from our simulation with a $3\mathrm{M_J}$ planet. In all locations where a non-Keplerian signature was indicated in the data (arrows), we find a counterpart in the model which is significant in residuals from a no planet model (Appendix~\ref{sec:noplanet}). Essentially a kink occurs every time the planet wake crosses the channel, as predicted analytically by \citet{Bollati:2021wl}. The close match between the predicted and actual kinematic signatures in the channel maps, while not proof, shows that the hypothesis of a single embedded planet can explain the widespread non-Keplerian signatures in the disc without recourse to large scale instabilities.

\subsection{Spiral arms in scattered light}
\label{sec:res_scattered}

Figure~\ref{fig:scattered_light} compares the VLT/SPHERE image of IM Lupi from \citet{Avenhaus:2018tb} (left panel) to the $\lambda=1.6\mathrm{\mu m}$ polarised intensity image from our $2\mathrm{M_J}$ simulation (right panel). Our model replicates the upper and lower disc surface shape, extending out to the same approximate radius. Most interesting is that we reproduce the inner spirals in the SPHERE data. In our model these are caused by the propagation of the planet wake in the upper layers of the disc. The simulated spirals from the planet extend to the edge of the emitting surface, as in the observations. The dark line along the upper-left diagonal from the centre of the model image is caused by the phase function of the polarisation \citep{Pinte:2006uj}. A drop in polarised intensity is also visible along the same diagonal in the observations.

\begin{figure*}
    \centering
    \includegraphics[width=\textwidth]{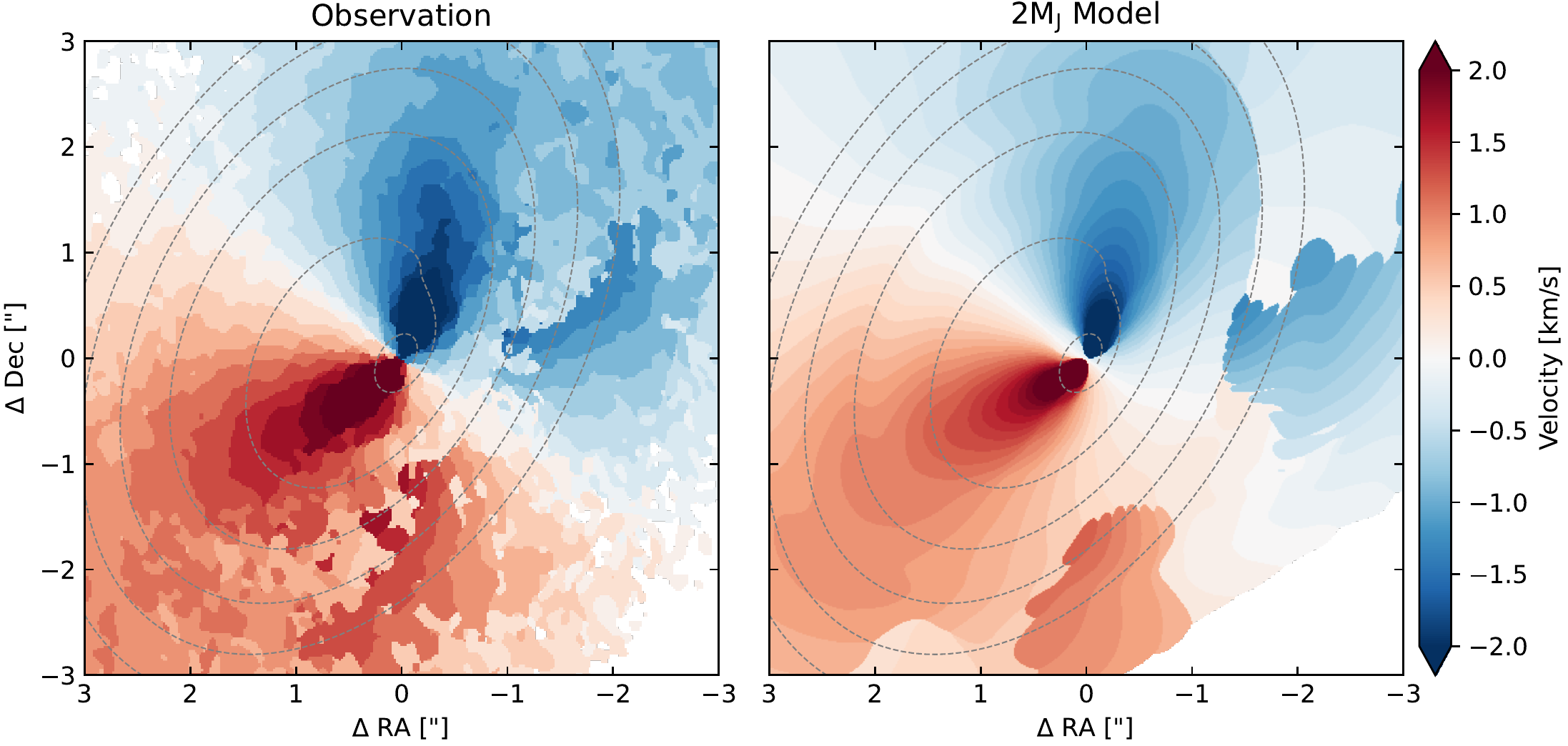}
    \vspace{-0.5cm}
    \caption{Left panel: peak velocity map in $^{12}$CO emission from MAPS \citep{Oberg:2021vx} data. Right panel: synthetic map from our $2\mathrm{M_J}$ model. The simulated $2\mathrm{M_J}$ result uses $\Delta v=0.2\mathrm{km/s}$. Dotted line (overlaid on both our simulation model and the observations) shows the expected location of the spiral wake, projected to the upper $^{12}$CO-emitting surface layer height derived by \citet{Law:2021vy}.}
    \label{fig:moment}
\end{figure*}

\subsection{Tracing the planet wake in the outer disc}
\label{sec:moment}

A key prediction from our model is that the planetary wake should produce coherent velocity perturbations tracing the wake in the outer disc. That is, the non-Keplerian features seen in Figure~1 should trace the specific morphology of a planet wake, as opposed to being a series of flocculent spiral arms from gravitational instability. \citet{Calcino:2021hp} found that the planet wake in the HD163296 disc was best traced in the peak velocity map.

Figure~\ref{fig:moment} compares the observed peak velocity map computed from the CO cube for IM Lupi supplied by the MAPS team (\citealt{Law:2021vy}; left) to same predicted from our $2\mathrm{M_J}$ model (right). To make the comparison clear, we overlaid both plots with a dotted line showing the analytic spiral wake \citep{Rafikov:2002rr} best matching our simulation model, projected to the upper disc surface using the $^{12}$CO emitting surface height calculated by \citet{Law:2021vy} (following \citealt{Pinte:2018uz}). In both the model and the observations, we observe velocity perturbations that trace the planet wake through the outer disc. The distorted contours in the observational data appear to follow the planet wake both inside and outside the planet location; particularly in the second spiral. Particularly intriguing is the `N-wave' structure apparent in the distorted contours around the dashed line in the observational map, which are predicted by both our simulations (right panel) and by semi-analytic models of planet wake propagation \citep{Rafikov:2002rr,Bollati:2021wl}.

Patches of red and blue seen in the bottom right of the data are simply where the lower surface of the disc becomes visible in the peak velocity map; this is reproduced in our model.

\begin{figure*}
    \centering
    \includegraphics[width=\textwidth]{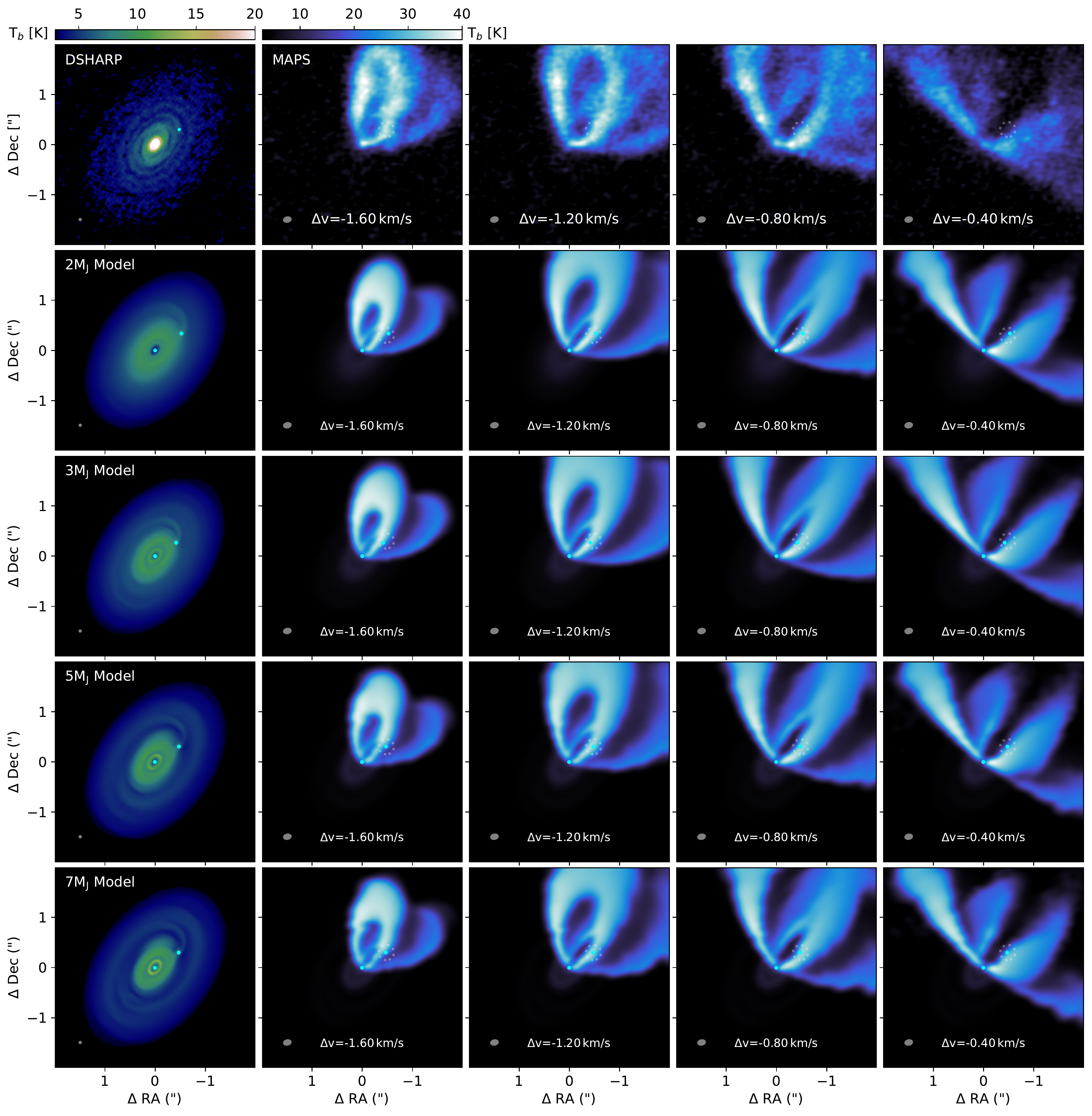}
    \caption{Effect of planet mass, comparing simulations (bottom rows) with observed dust continuum emission (left column) and selected $^{12}$CO channels (right panels). Blue dot shows the expected planet position while the grey circle indicates predicted location of the planet at 117~au \citep{Pinte:2020vw}. Filled ellipse gives beam size.  Radiative transfer images and cubes are available on FigShare:\dataset[doi:10.26180/20145620.v3]{https://doi.org/10.26180/20145620.v3}}
    \label{fig:IM_Lup_plots}
\end{figure*}

\subsection{Continuum emission and planet mass estimate}

Figure~\ref{fig:IM_Lup_plots} shows the continuum emission (left column) and selected CO channels (right columns) from each of the four simulations to data from DSHARP and MAPS \citep{Andrews:2020wh,Czekala:2021wt} (top row).

The kink observed by \citet{Pinte:2020vw} is located in the $\Delta v=-1.60\mathrm{km/s}$ channel, which is also seen in our simulations. %Furthermore, dark bands in the $\Delta v=-1.60$ km/s and $\Delta v=-0.80$ km/s slices are also reproduced by the simulations. 
Some additional perturbations (or `secondary kinks') can be seen on both the upper and lower surfaces, as in Figure~1. These correspond to velocity perturbations created by the planet wake.

Comparing the DSHARP continuum (\citealt{Andrews:2020wh}; top panel) to the four simulations (bottom four panels), on all simulation models we observe an $m=2$ spiral in the disc, increasing in amplitude with planet mass. For larger planet masses the simulations also show the formation of a gap in the outer ring, carved by the planet. However for planet masses $\geq 5$M$_{\rm Jup}$ the amplitude of the scattered light spirals are too high compared to the data (Figure~\ref{fig:scattered_light}), so we favour a planet mass of 2--3 M$_{\rm Jup}$.

%%%%%%%%%%%%%%%%%%%%%%%%%%%%%%%%%%%%%%%%%%%%%%%
%%%%%%%%%%%%%%%%% DISCUSSION %%%%%%%%%%%%%%%%%%

\section{Discussion}
\label{sec:discussion}
In this Letter we explored whether an embedded planet can explain the spiral arms, non-Keplerian motions and scattered light emission in the IM Lupi disc. With caveats, we find that a single planet of several times the mass of Jupiter orbiting at $\approx~100$--$120$~au can simultaneously explain all of these features. Our best evidence for a planet is in the kinematics. We confirm that the tentative kink found by \citet{Pinte:2020vw} in the new data from MAPS \citep{Oberg:2021vx}. However this is not the only kink in the data. In particular, a series of `secondary kinks' \citep{Bollati:2021wl} appear to trace the planet wake in the observations in a manner predicted by our simulations. These are best seen in the observed peak velocity map (Figure~\ref{fig:moment}), as recently demonstrated for the planet in the HD163296 disc \citep{Pinte:2018va, Calcino:2021hp}.

To reproduce the disc morphology in scattered light (with or without a planet) we needed the grains responsible for scattered light emission to be well-coupled to the gas, and for the disc to be optically thick at 1.6$\mu$m. In our hydrodynamic model we simulated a range of grain `sizes' --- assuming spherical grains and an intrinsic grain density of 3 ${\rm g}/{\rm cm}^3$ --- from 1$\mu$m to 2.3~mm, corresponding to Stokes numbers of $10^{-4}$ and $0.2$ at 100~au, respectively (e.g. using Eq.~3 in \citealt{Dipierro:2015ty}). In the radiative transfer calculation we needed an additional factor of ten scaling (down) in the Stokes number to avoid settling of the grains responsible for the scattered light emission out of the disc atmosphere. This is not surprising as \citet{Cleeves:2016ux} already inferred that micron-sized grains have the same scale height as the gas in IM Lupi ($h=12$au at $r=100$au) and that the scale height of mm-emitting grains is $\approx 4 \times$ smaller. This likely indicates that grains are fluffy aggregates rather than compact spheres, as already suggested for IM Lupi by \citet{Pinte:2008pm} and by ALMA polarization observations \citep{Hull:2018yz}.

Elias 2-27 is the other main disc that shows $m=2$ spiral arms in continuum emission \citep{Huang:2018}. The hypotheses are a planet, or gravitational instability \citep{Meru:2017ji,Forgan:2018if}. Gravitational instability can lead to formation of multiple spiral arms and possibly fragmentation to form gas giants \citep{KimuraShigeoS2012CoGI,KratterKaitlin2016GIiC} and hence may also explain the main phenomena seen in IM Lupi. But one would need to posit that the inner disc is unstable (to produce spiral arms) while the outer disc, being relatively featureless in continuum, is not. This is opposite to the expected scenario where the outer disc is more unstable \citep{KratterKaitlin2016GIiC}. The global condition for instability is ${\rm M}_{\rm disc}/{\rm M}_* \gtrsim H/R$ \citep{Toomre:1964uo}. With ${\rm M}_{\rm disc}=0.1$~M$_\odot$ the disc-to-star mass ratio is $0.1/1.12=0.09$ which is smaller than the inferred $H/R\approx~0.12$ at 100~au and hence suggests that the disc is stable. \citet{Sierra:2021wg} estimated a Toomre $Q\approx$2 at $r\gtrsim~50$~au in IM Lupi, suggesting the disc is gravitationally stable. We neglected self-gravity in our models but the above discussion suggests that it could be important in IM Lupi, even if not ultimately responsible for the spiral arms.

%Gravitational instability is difficult to model directly for a particular source because one needs to account for both the irradiation from the central star and the internal disc heating from shocks, although the approach taken by \citet{Meru:2017ji} to fix the surface temperature profile to mimic irradiation is one possible approach.

%Our simulations created distinct spiral arms; structures that have been observed in both the dust continuum and scattered light images captured by DSHARP and SPHERE respectively. These spirals are more prominent in the larger mass simulations. The gaps in the disc seen in the dust continuum were not completely replicated by our models, but this could be due to the short simulation time used. $8.7 \times 10^3$ years is not a representative time-frame for disc evolution, with the average lifetime of a protoplanetary disc to be a couple million years \citep{Baillie:2019vw}. We expect the gap that starts to form in the $5\mathrm{M_J}$ simulation to increase in size with more orbits. Additionally, the spiral arms shown in the scattered light image are replicated in smaller planet simulations. The $2\mathrm{M_J}$ demonstrates that smaller bodies are still able to produce spirals that can be seen in observations. Figure~\ref{fig:scattered_light} showcases the effect that a smaller planet can make on the surface spirals.

Planets in protoplanetary discs migrate \citep{Ward:1986aa}. In {\sc Phantom}, sink particles are free to interact with gas and dust, and hence migrate. We found that the $5\mathrm{M_J}$ planet migrated at $\approx 2$ au/kyr after gap opening. Additionally, the migration depends on the planet mass, with more massive planets migrating faster. This can be seen in Figure~\ref{fig:IM_Lup_plots} by comparing the grey circles (indicating the 117~au estimate) and the simulated location. Migration --- which can be inward or outward depending on disc properties --- makes it difficult to predict the radial location of the planet and brings a coincidence problem as to why any planet would be observed at a particular location.
Nevertheless, we have shown that a planet in the right location can explain all the main substructures observed in IM Lupi.

%%%%%%%%%%%%%%%%%%%%%%%%%%%%%%%%%%%%%%%%%%%%%%%
%%%%%%%%%%%%%%%%% CONCLUSION %%%%%%%%%%%%%%%%%%

\section{Conclusions}
\label{sec:conclusion}
%Our main findings are:
\begin{enumerate}
\item A single embedded 2--3~M$_{\rm Jup}$ planet orbiting at $\approx$110~au in the IM Lupi circumstellar disc can simultaneously explain the 16 different localised deviations from Keplerian motion (`kinks') seen in the $^{12}$CO channel maps.
\item The same planet can explain the spiral arms seen in the upper layers of the disc in scattered light and the $m=2$ spiral arms seen in continuum emission.
\item We predict, and confirm, that the wake from the planet should be visible in the observational peak velocity map from the $^{12}$CO emission line. The perturbations seen in this map (Fig.~\ref{fig:moment}; left panel) closely match the prediction for the planet-generated spiral arm
\end{enumerate}

We required a relatively massive disc ($\approx~0.1$M$_\odot$) as well as a scaled Stokes number for the dust grains to remain well-coupled in both the upper layers and midplane of the disc. Such a disc mass suggests that gravitational instability may also be possible (but with \citealt{Sierra:2021wg} finding $Q_{\rm min}\sim~2$).

\section*{Acknowledgments}
DJP thanks Judit Szulagyi for useful discussions about the perils of vacuum cleaner sink particles at the `Kinematics of Planet Formation' meeting in 2019; also Giuseppe Lodato and Richard Teague for useful discussions. JC acknowledges support from LANL/LDRD program (approved for release as LA-UR-21-31387). We used OzSTAR and Gadi supercomputing facilities funded by Swinburne University (OzSTAR) and the Australian Government via the National Computing Initiative. DP and CP acknowledge Australian Research Council funding via DP180104235. We made use of ALMA data: ADS/JAO.ALMA\#2018.1.01055.L, ADS/JAO.ALMA\#2016.1.00484.L. ALMA is a partnership of ESO (representing its member states), NSF (USA) and NINS (Japan), together with NRC (Canada), MOST and ASIAA (Taiwan), and KASI (Republic of Korea), in cooperation with the Republic of Chile. We thank the referee for useful suggestions.

\appendix

\section{Residuals from Keplerian rotation}
\label{sec:noplanet}
\begin{figure*}
    \centering
    \includegraphics[width=\textwidth]{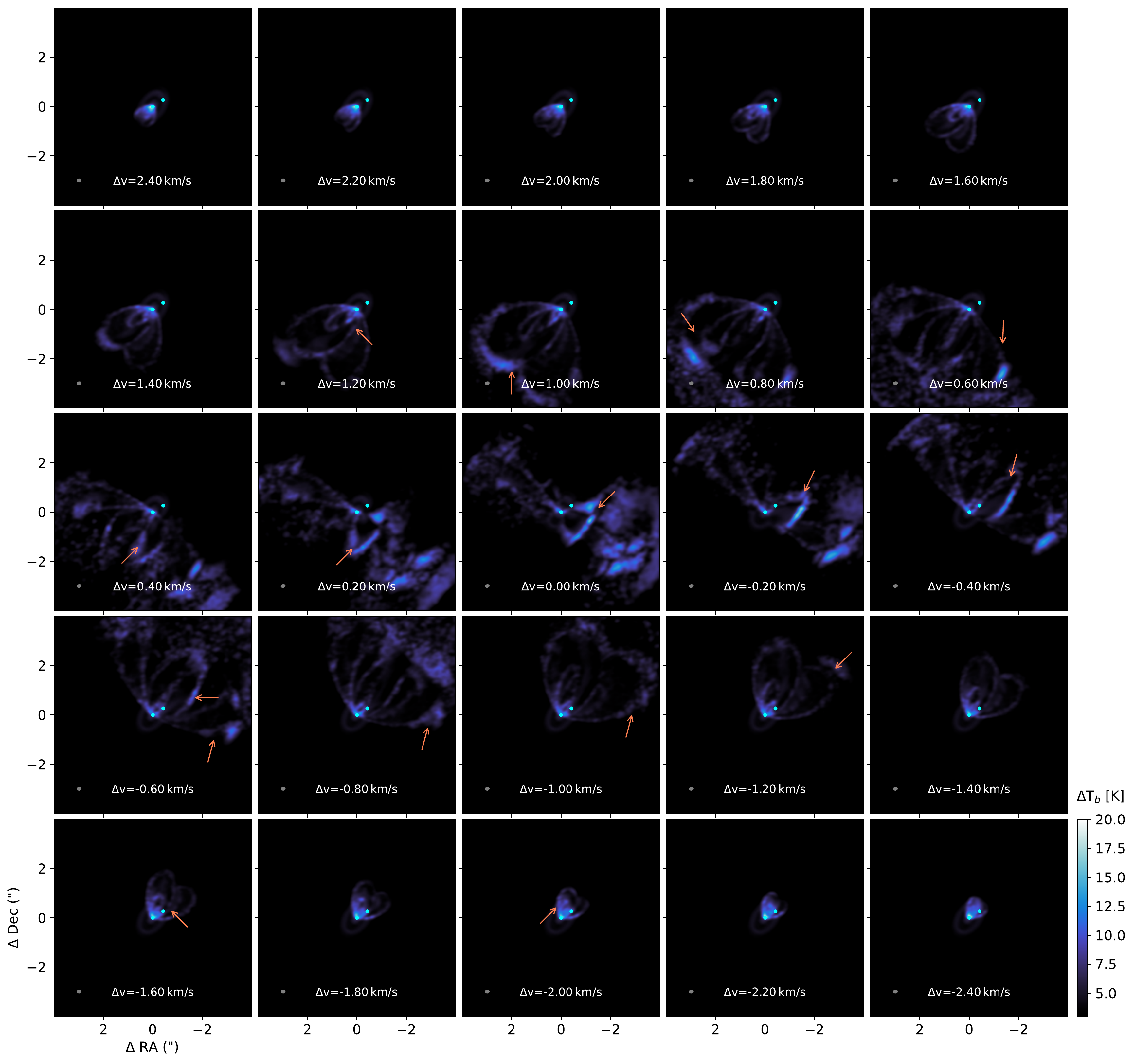}
    \caption{Absolute brightness temperature residuals in channel maps of $^{12}$CO J=2--1 line emission after subtracting a model with no planet from the model shown in Figure~\ref{fig:channels_model}, highlighting significant non-Keplerian features in individual channels caused by the embedded planet. Arrows are as identified from observations in Figure~\ref{fig:channels}.}
    \label{fig:residuals}
\end{figure*}

To assess the significance of the non-Keplerian features in the channel maps presented in Figures~\ref{fig:channels} and \ref{fig:channels_model}, Figure~\ref{fig:residuals} shows the absolute brightness temperature residuals in the channel maps when the model with no planet is subtracted from the model shown in Figure~\ref{fig:channels_model}. Arrows reproduced from Figures~\ref{fig:channels}--\ref{fig:channels_model} all point to correspondingly significant features ($\Delta T_b > 10$ K; corresponding to a signal-to-noise ratio of $\gtrsim 7$) in the residual maps. We found similar results when subtracting an azimuthally averaged model.
%%%%%%%%%%%%%%%%%%%% REFERENCES %%%%%%%%%%%%%%%%%%

\label{lastpage}
\end{document}